\documentclass{emulateapj}
\usepackage{natbib}
\begin{document}
\bibliographystyle{prsty}

\title{Universal Nonlinear Small-Scale Dynamo}
\author{A. Beresnyak}
\affiliation{Los Alamos National Laboratory, Los Alamos, NM, 87545}
\affiliation{Ruhr-Universit\"at Bochum, 44780 Bochum, Germany}

\begin{abstract}
  We consider astrophysically relevant nonlinear MHD dynamo at large Reynolds numbers (Re).
  We argue that it is universal
  in a sense that magnetic energy grows at a rate which
  is a constant fraction $C_E$ of the total turbulent dissipation rate.
  On the basis of locality bounds we claim
  that this ``efficiency of small-scale dynamo``, $C_E$, is a true constant
  for large Re and is determined only by strongly nonlinear dynamics at the equipartition scale.
  We measured $C_E$ in numerical simulations and 
  observed a value around $0.05$ in highest resolution simulations.
  We address the issue of $C_E$ being small, unlike Kolmogorov constant
  which is of order unity.

\end{abstract}

\maketitle

\section{Introduction} 

MHD turbulence is ubiquitous in astrophysical and space
environments  \citep{goldstein1995,armstrong1995}. Reynolds numbers are, typically, very high, owing
to large scales of astrophysical processes compared to small dissipative scales.
One of the central questions of MHD dynamics is how
initially unmagnetized well-conductive fluid generates its own magnetic field, known broadly as ``dynamo''.
Turbulent dynamo has been roughly subdivided into ``large-scale dynamo'' and ``small-scale'' or ``fluctuation dynamo'' depending
on whether magnetic fields are amplified on scales larger or smaller than turbulence outer scales.
Although a few ``no-dynamo'' theorems has been proved for flows with symmetries, a generic
turbulent flow, which possess no exact symmetries was expected to amplify
magnetic fields by stretching, due to particle separation in a turbulent flow.
For the large-scale field, a ``twist-stretch-fold'' mechanism was
introduced   \citep{vainshtein1972}. Turbulent flow possessing perfect
statistical isotropy can not generate large-scale magnetic field. Observed large-scale fields,
such as in disk galaxies, are generated when statistical symmetries of turbulence are
broken by large-scale asymmetries of the system, such as stratification, rotation
and shear  \citep[see, e.g.,][]{vishniac2001,kapyla2009}. Since these symmetries are only
weakly broken, large-scale dynamo is slow. Small-scale dynamo does not suffer from this
restriction and can be fast. Kinematic small-scale dynamo, which ignores the backreaction of the magnetic
field has been studied extensively before   \citep{kazantsev1968,kulsrud1992}. However, kinematic
dynamo is irrelevant for astrophysical environments, because it grows exponentially and becomes
inapplicable at smallest turbulent timescales which are tiny by astrophysical standards. 
Additionally, kinematic dynamo have positive spectral index (typically 3/2) at all scales,
which is incompatible with observations in galaxy clusters  \citep{Laing2008} which clearly
indicate steep spectrum with negative power index at small scales. Due to preexisting
astrophysical fields small-scale dynamo starts in nonlinear regime.
It was discovered numerically that small-scale dynamo continues to grow after kinematic
stage, producing steep spectrum at small scales and significant outer-scale fields  \citep{haugen2004,brandenburg2005,CVB09,ryu2008}.
Furthermore, MHD turbulence produces turbulent diffusivity
(aka ``$\beta$-effect``), which is essential for large-scale
dynamo  \citep{kapyla2009} and reconnection  \citep{lazarian1999,eyink2010}.
Saturation of small-scale dynamo seems to be independent on $Re$ and $Pr$ as long as $Re$ is large \citep{haugen2004}
and the magnetic energy growth rate could be constant \citep{schekochihin2007, CVB09, BJL09, ryu2008}.
Small-scale dynamo is faster than large-scale dynamo in most astrophysical
environments and magnetic energy
grows quickly to equipartition with kinetic motions, with the largest scales of such field
being a fraction of the outer scale of turbulence. Subsequently, these turbulent fields are slowly
ordered by mean-field dynamo, with turbulent diffusivity of MHD turbulence playing essential role. 
In this paper we provide sufficient analytical and numerical argumentation behind the universality
of nonlinear small-scale dynamo.

\section{Nonlinear small-scale dynamo}

\begin{figure}
\includegraphics[width=0.8\columnwidth]{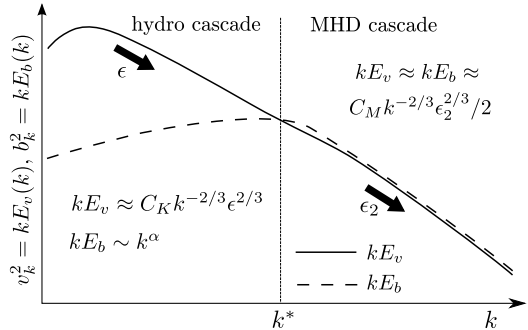}
\caption{A cartoon of kinetic and magnetic spectra in small-scale dynamo, at a particular moment of
time when equipartition wavenumber is $k^*$.}
\label{spect_dyn}
\end{figure}

We assume that the spectra of magnetic and kinetic energy
at a particular moment of time are similar to what is presented on Fig.~\ref{spect_dyn}.
Magnetic and kinetic spectra cross at some ``equipartition'' scale $1/k^*$,
below which both spectra are steep due to MHD cascade~ \citep[see, e.g.,][]{GS95,B11}. This is suggested
by both numerical evidence \citep{BL09b,CVB09}
observations of magnetic fields in clusters of galaxies \citep{Laing2008}. At larger scales magnetic spectrum is shallow, $k^\alpha,\, \alpha>0$,
while kinetic spectrum is steep due to a hydro cascade. Most
of the magnetic energy is concentrated at scale $1/k^*$. We designate $C_K$ and $C_M$ as Kolmogorov constants
of hydro and MHD respectively. The hydrodynamic cascade rate is $\epsilon$ and the MHD cascade rate as $\epsilon_2$.
Due to the conservation of energy in the inertial range, magnetic energy will grow at a rate $\epsilon-\epsilon_2$.
We will designate $C_E=(\epsilon-\epsilon_2)/\epsilon$ as an ``efficiency of small-scale dynamo'' and
we will argue that this is a true constant, since: a) turbulent dynamics is local in scale in
the inertial range; b) neither ideal MHD nor Euler equations contain any scale explicitly.
Magnetic energy, therefore, grows linearly with time if $\epsilon=const$. 
The equipartition scale $1/k^*$ will grow with time as $t^{3/2}$ \citep{BJL09}. This
is equivalent to saying that small-scale dynamo saturates at several dynamical times
at scale $1/k^*$ and proceeds to a twice larger scale \citep{schekochihin2007}. If magnetic
energy grows approximately till equipartition \citep{haugen2004,CVB09}, the whole process
will take around several dynamical timescales of the system, or more quantitatively, $(C_K^{3/2}/C_E)(L/v_L)$.

\section{Locality of small-scale dynamo}

We will use ``smooth filtering'' approach with
dyadic-wide filter in k-space \citep{aluie2010,eyink2005}. We designate a filtered vector quantity
as ${\bf a}^{[k]}$ where $k$ is a center of a dyadic Fourier filter in the range of wave numbers
$[k/2,2k]$. The actual logarithmic width of this filter is irrelevant to further
argumentation, as long as it is not very small. We will assume that the vector field ${\bf a}$ is
H\"olger-continuous with some exponent and designate $a_k=\langle|{\bf a}^{[k]}|^3\rangle^{1/3}$
which has to scale as $k^{\sigma_3}$, e.g., $k^{-1/3}$ for velocity in Kolmogorov
turbulence. The energy cascade rate is $\epsilon=C_K^{-3/2}kv_k^3$, where here 
we defined Kolmogorov constant $C_K$ by third order, rather than second order quantities.
We will keep this designation, assuming that traditional Kolmogorov constant could be used instead.
We use spectral shell energy transfer functions such as
$T_{vv}(p,k)=-\langle{\bf v}^{[k]}({\bf v}\cdot{\bf \nabla}){\bf v}^{[p]}\rangle, \,
T_{w^+w^+}(p,k)=-\langle{\bf w^+}^{[k]}({\bf w^-}\cdot{\bf \nabla}){\bf w^+}^{[p]}\rangle$
 \citep{alexakis2005}, applicable to incompressible ideal MHD equations,
where $w^\pm$ are Els\"asser variables and $v$, $b$ and $w^\pm$ are measured in the same Alfvenic units.
We will use central frequency $k$ and study ``infrared`` (IR) transfers
from $p\ll k$, and ''ultraviolet'' (UV) transfers, from $q\gg k$.
We will provide bounds on $|T|$, in units of energy rate as in \citet{aluie2010,eyink2005}
and relative volume-averaged bounds which are divided the actual energy rate and are dimensionless.
We will consider three main
wavevector intervals presented on Fig.~\ref{spect_dyn}: $k\ll k^*$ (``hydro cascade''), $k\sim k^*$ (``dynamo'')
and $k \gg k^*$ (``MHD cascade'').

\section{MHD cascade, $k \gg k^*$.}

The only energy cascades here are Els\"asser cascades and, by design of
our problem, $w^+$ and $w^-$ have the same statistics, so we will drop $\pm$.
For an exchange with $p\ll k$ band, for $|T_{ww}|$, using H\"olger inequality and wavenumber
conservation we get an upper bound of $pw_pw_k^2$ and for $q \gg k$ band it is $kw_q^2w_k$,
these bounds are asymptotically small, see \citet{eyink2005}. For the full list of transfers and
limits refer to Table~\ref{transfers}.
The relative bound should be taken with respect to
$C_M^{-3/2}kw_k^3$, where $C_M$ is a Kolmogorov constant for MHD,
from which we get that most of the energy transfer with the $[k]$ band should
come from $[kC_M^{-9/4},kC_M^{9/4}]$ band, see \citet{B11}.
The global transfers between kinetic and magnetic energy must average out in this regime,
nevertheless, the pointwise IR and UV transfers can be bounded by $pb_pv_kb_k$ and $kb_q^2v_k$
and are small \citep{eyink2005}.

\begin{table}[t]
\begin{center}
\caption{Transfers and upper limits}
  \begin{tabular*}{0.99\columnwidth}{@{\extracolsep{\fill}}r c l c c}
    \hline\hline

Transfers & & & $p\ll k$ & $q\gg k$ \\

   \hline

$T_{vv}(p,k)$&=&$-\langle{\bf v}^{[k]}({\bf v}\cdot{\bf \nabla}){\bf v}^{[p]}\rangle$
 & $pv_pv_k^2$ & $kv_kv_q^2$  \\

$T_{bb}(p,k)$&=&$-\langle{\bf b}^{[k]}({\bf v}\cdot{\bf \nabla}){\bf b}^{[p]}\rangle$
 &  $pv_pv_k b_k$ & $kb_kv_qb_q$ \\

$T_{vb}(p,k)$&=&$\langle{\bf b}^{[k]}({\bf b}\cdot{\bf \nabla}){\bf v}^{[p]}\rangle$
 &  $pv_pb_k^2$ & $kb_kv_qb_q$  \\

$T_{bv}(p,k)$&=&$\langle{\bf v}^{[k]}({\bf b}\cdot{\bf \nabla}){\bf b}^{[p]}\rangle$
 &  $pb_pv_k b_k$ & $kv_kb_q^2$  \\

$T_{w^+w^+}(p,k)$&=&$-\langle{\bf w^+}^{[k]}({\bf w^-}\cdot{\bf \nabla}){\bf w^+}^{[p]}\rangle$
 &  $pw_pw_k^2$ & $kw_kw_q^2$  \\

   \hline

\end{tabular*}
  \label{transfers}
\end{center}
\end{table}

\section{Hydro cascade, $k\ll k^*$}

Despite having some magnetic energy
at these scales, most of the energy transfer is dominated by velocity field.
Indeed, $|T_{vv}|$ is bounded by $pv_pv_k^2$ for $p \ll k$ and by $kv_q^2v_k$ for $q \gg k$. Compared to these,
$|T_{bv}|$ transfers are negligible: $pb_pv_kb_k$ and $kb_q^2v_k$. For magnetic energy in
$p\ll k$ case we have $|T_{vb}|$ and $|T_{bb}|$ transfers bounded by $pv_pb_k^2$, $pb_pv_kb_k$ and for
$q\gg k$ case $|T_{vb}|$ and $|T_{bb}|$ are bounded by $kb_kv_qb_q$.
Out of these three expressions the first two go to zero, while the third goes
to zero if $\alpha-2/3<0$ or have a maximum at $q=k^*$ if $\alpha-2/3>0$.
This means that for the transfer to magnetic energy we have IR locality, but not necessarily
UV locality. Note that magnetic energy for $k\ll k^*$ is small
compared to the total, which is dominated by $k=k^*$.
We will assume that $\alpha-2/3>0$ and that the spectrum of $b_k$ for $k<k^*$ is formed
by nonlocal $|T_{vb}|$ and $|T_{bb}|$ transfers from $k^*$, namely magnetic structures at $k$ are formed
by stretching of magnetic field at $k^*$ by velocity field at $k$. Magnetic spectrum
before $k^*$ is, therefore, nonlocal and might not be a power-law.

\section{Dynamo cascade $k=k^*$}

In this transitional regime our estimates
of Els\"asser UV transfer and kinetic IR transfer from two previous sections will
hold. We are interested how these two are coupled together and produce
observable magnetic energy growth or decay.
IR $p\ll k^*$ $|T_{vb}|$ and $|T_{bb}|$ transfers will be bounded by $pv_pb_{k^*}^2$ and
$pb_pv_{k^*}b_{k^*}$,
which go to zero, so there is a good IR locality. Ultraviolet transfers
will be bounded by $k^*b_{k^*}b_qv_q$. This quantity also goes to
zero as $q$ increases, so there is an UV locality for this regime as well.
Let us come up with bounds of relative locality. Indeed, the actual growth of
magnetic energy was defined as $\epsilon_B=\epsilon-\epsilon_2=C_EC_K^{-3/2}kv_k^3$.
So, $p\ll k^*$ IR bound is $k^*C_E^{3/2}C_K^{-9/4}$
and UV bound is $k^*C_E^{-3/2}C_M^{9/4}$. We conclude that most of the interaction
which result in magnetic energy growth must reside in the wavevector interval of 
$k^*[C_E^{3/2}C_K^{-9/4},C_E^{-3/2}C_M^{9/4}]$. Numerically, if we substitute
$C_K=1.6$, $C_M=4.2$, $C_E=0.05$ we get the interval of $k^*[0.004,2000]$.
So, despite being asymptotically local, small-scale dynamo can be fairly nonlocal from
practical standpoint.

\begin{table}[t]
\begin{center}
\caption{Three-dimensional MHD simulations}
  \begin{tabular*}{0.99\columnwidth}{@{\extracolsep{\fill}}c c c c c c c}
    \hline\hline
Run  & n & $N^3$ & Dissipation & $\langle\epsilon\rangle$ &  Re & $C_E$ \\

   \hline

M1-6 & 6 & $256^3$   &  $-7.6\cdot10^{-4}k^2$  & 0.091 & 1000 & $0.031\pm0.002$ \\

M7-9 & 3 &  $512^3$  &  $-3.0\cdot10^{-4}k^2$  & 0.091 & 2600 & $0.034\pm0.004$  \\

M10-12 & 3 & $1024^3$  &  $-1.2\cdot10^{-4}k^2$ & 0.091 & 6600  & $0.041\pm0.005$ \\

M13 & 1 & $1024^3$ & $-1.6\cdot10^{-9}k^4$   & 0.182 & -- & $0.05\pm0.005$  \\

M14 & 1 & $1536^3$ & $-1.5\cdot10^{-15}k^6$  & 0.24  & -- & $0.05\pm0.005$ \\

   \hline

\end{tabular*}
  \label{experiments}
\end{center}
\end{table}


To summarize, kinetic cascade at large scales and MHD cascade at small scales
in the inertial range are dominated by local interactions. The transition between kinetic
cascade and MHD cascade is also dominated by local interactions, and since ideal MHD equations
do not contain any scale explicitly, the efficiency of small-scale dynamo $C_E$ is a true universal
constant.
Note that $C_E$ relates energy fluxes, not energies, so this claim is unaffected by
the presence of intermittency. Magnetic spectrum at $k\ll k^*$ is dominated by nonlocal
triads that reprocess magnetic energy from $k= k^*$ and, since this
part of the spectrum contains negligible magnetic energy, our
universality claim is unaffected by this nonlocality.

\section{Numerical results}

\begin{figure}
\includegraphics[width=0.8\columnwidth]{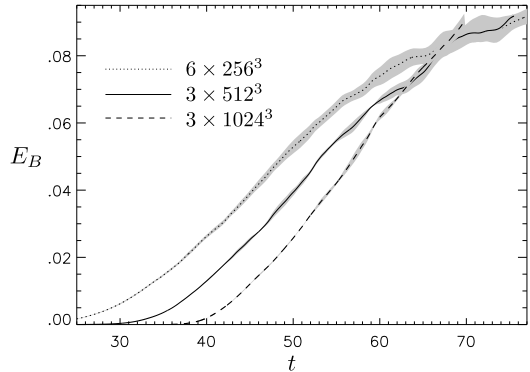}
\caption{Magnetic energy growth observed in simulations M1-6, M7-9 and M10-13. We used sample averages which greatly
reduced fluctuations and allowed us to measure $C_E$ with sufficient precision.}
\label{dyn1}
\end{figure}

We performed numerical simulations of statistically homogeneous isotropic
small-scale dynamo by solving MHD equations with stochastic non-helical driving and explicit
dissipation with $Pr_m=1$.
The details of the code and driving are described in detail in our earlier publications  \citep{B11, BL09a} and 
Table~1 shows simulation parameters. We started each simulation from previously well-evolved driven
hydro simulation by seeding low level white noise magnetic field. We ran
several statistically independent simulations in each group and obtained growth rates and errors
from statistical (or sample) averages. In all simulations, except M14, the energy injection rate was controlled.
Fig~\ref{dyn1} shows sample-averaged time evolution of magnetic energy. Growth is initially exponential
and smoothly transition into the linear stage. Note, that scatter is initially small, but grows
with time, which is
consistent with the picture of magnetic field growing at progressively larger scales and
having progressively less independent realizations in a single datacube.

\section{Efficiency of small-scale dynamo}

Our $C_E$ is much
smaller than unity. One would expect a quantity of order unity because this is
a universal number, determined only by strong interaction on equipartition scale.
If we refer to the ideal incompressible MHD equations, written in terms of Els\"asser variables,
$\partial_t{\bf w^\pm}+\hat S ({\bf w^\mp}\cdot\nabla){\bf w^\pm}=0$, dynamo could be understood
as decorrelation of ${\bf w^\pm}$ which are originally perfectly correlated with ${\bf w^+}={\bf w^-}$
in the hydrodynamic cascade. In our case this decorrelation is happening at the equipartition scale $k^*$.
Being time-dependent, this decorrelation propagates upscale, while ordinary energy cascade goes downscale.
The small value of $C_E$ might be due to this.
A different viewpoint on small-scale dynamo is a picture when magnetic field is being
stretched by kinetic motions and, therefore, amplified. With small or negligible diffusion the field
becomes extremely tangled \citep{Schekochihin2004}, quite contrary to the observations and numerical
experiments, where magnetic field has an energy-containing scale which is larger than the scale
of resistive dissipation. In the nonlinear regime, therefore, the intuitive picture
with stretching with resistive diffusion is completely wrong. It is inevitable to appeal
to turbulent diffusion which helps to create larger-scale field. Both stretching and diffusion
depend on turbulence properties at the same designated scale $1/k^*$, therefore
in the asymptotic regime of large $Re$ one of these processes must dominate. As we know
$C_E$ is small so stretching and diffusion are close to canceling each other.

\section{Kinematic dynamo rates}

A better studied and understood kinematic dynamo
might shed some light on the problem of small $C_E$. In the kinematic regime,
when we neglect Lorentz force in the MHD equation, the growth is exponential and the rate is expected
to come from fastest shearing rates of smallest turbulent eddies. Observed rates, however, are smaller
which was interpreted as competition between stretching and turbulent mixing \citep{eyink2011}.
In our simulations, in the kinematic regime of M7-9 we observed growth rate $\gamma\tau_\eta=0.0326$,
where $\tau_\eta=(\nu/\epsilon)^{1/2}$ is a Kolmogorov timescale,
which is consistent with  \citep{haugen2004,Schekochihin2004}. With minimum timescale,
$\tau_{\rm min}\approx 9\tau_\eta$, $\gamma\tau_{\rm min}=0.3$, which is still small.

Kazantsev-Kraichnan model \citep{kazantsev1968,*kulsrud1992} 
predicts $\gamma\tau_{\rm min}\sim 1$. This model, however, uses ad-hoc delta-correlated velocity
which does not correspond to any dynamic turbulence and its statistics is time-reversible
as opposed to time-irreversible real turbulence. Time irreversibility of hydro turbulence
actually mandates that fluid particles separate faster backwards in time, since
$\langle v_{\|l}^3 \rangle=-4/5\epsilon l$ is negative.

\begin{figure}
\includegraphics[width=0.8\columnwidth]{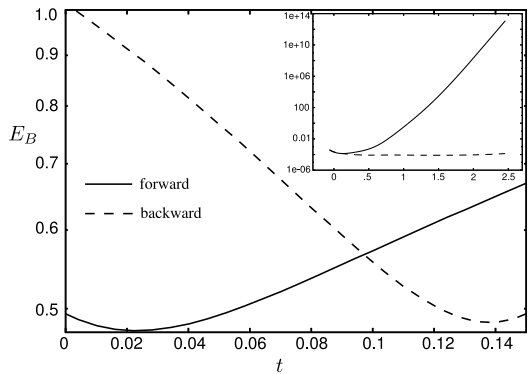}
\caption{Evolution of magnetic energy forward and backward in time (dashed is with reversed x-axis).
Inset: a naive simulation with initial state $+{\bf v}$ and $-{\bf v}$ from Navier-Stokes simulation
(not a rigorous backward in time simulation).}
\label{backw}
\end{figure}

In order to study the interplay of stretching and diffusion, we performed several simulations of
kinematic dynamo forward and backward in time. We followed full three-dimensional evolution
of v and b and approximated ``backward in time'' by reversing velocity direction.
Initial condition for magnetic field was typically random noise.
Since we couldn't reverse viscous losses in DNS, we used viscosity $\nu=0$, but magnetic diffusivity $\eta>0$.
In the first set of simulations we set initial velocity as $v$ and $-v$ from evolved viscous runs.
The growth rates is shown on inset of Fig.~\ref{backw}. Quite surprisingly, the ``backward'' simulation
did not produce any growth for several dynamical times. Unexpectedly, simply
reversing velocity has such a profound impact on kinematic dynamo, despite spectra being very
close to each other, suggesting that it is not only the spectrum that determines growth but rather
the actual statistical properties of velocity, which will determine whether
stretching or diffusion wins, i.e. if there is a dynamo or a no-dynamo, even in the simple
kinematic case. In this simulation we observed a typical $k^2$ ``thermal pool`` at the
end of velocity spectrum which had shortest timescales. ''Thermal pool`` was clearly
time-irreversible, unlike the true physical thermal pool, consistent with \citep{ray2011}.

The next series of simulations were reproducing an actual backward in time dynamics.
In order to achieve this we evolved initial state for a fairly short time with $\nu=0$ and then
we evolved it for the same time reverting velocity with $\nu=0$ and confirmed that final state
is close to initial state, due to reversibility of truncated Euler equations. The results for dynamo
growth is shown of Fig.~\ref{backw}. We see that backward dynamo is faster by a factor of $2.0\pm0.1$,
which is actually consistent with the ratio of particle diffusion forward and backward in time in \citep{eyink2010}.
This result again reinforces our statement that dynamo is a result of competing mechanisms of turbulent
stretching and turbulent diffusion and the outcome depend on statistics of velocity
other than just velocity spectrum.


\section{Discussion}

Linear growth regime that we considered in this Letter provides a fastest growth rate for magnetic energy, $C_E\epsilon$,
which is actually just a constant fraction of absolute energetic limit of $\epsilon$. This linear growth starts with
$t=0$ in astrophysical enviroments with very high Re, as long as turbulence is well-developed. Exponential growth
regime, that is relevant for finite Re and small initial seed fields provide much slower growth. Combined with finite
time that is available in simulated astrophysical objects this may result in severe underestimation of the magnetic energy
and, therefore, of dynamical importance of magnetic fields. For example, cosmological simulations
of halo collapse, that have rather modest Re,
could be underestimating magnetic energy \citep{sur}. A resolution study must confirm
convergence of magnetic energy with resolution. If exponential growth is observed, magnetic energy is underestimated.

In this Letter we discussed the issue of $C_E$ being small and interpreted this as a competition between turbulent
stretching that enhances magnetic field and turbulent diffusion that tries to dissipate it. Indeed, the negative
values of $C_E$, from theoretical viewpoint, are also allowed. This would be a situation of non-dynamo
in the nonlinear regime.
Isotropic homogeneous Kolmogorov turbulence does produce a growth and most astrophysical enviroments
that are well-magnetized. However, turbulent magnetic diffusion can dominate in situations when
turbulence is very different from isotropic and homogeneous and have finite Re.
This might explain why dynamo experiments sometimes generate larger fields in the absense
of turbulence \citep{colgate}.


I am deeply grateful to Greg Eyink for extended discussions.
I am grateful to Hussein Aluie, Axel Brandenburg, Gregory Falkovich, Alex Lazarian,
Hui Li and Hao Xu for discussions.
I was supported by LANL Director's Fellowship and Humboldt Fellowship.
Computations were performed on Ranger through NSF TeraGrid allocation TG-AST080005N.



\begin{thebibliography}{10}

\bibitem[Alexakis et al(2005)]{alexakis2005}
A. {Alexakis}, P.~D. {Mininni}, and A. {Pouquet}, \pre {\bf 72},  046301
  (2005).

\bibitem[Aluie \& Eyink(2010)]{aluie2010}
H. {Aluie} and G.~L. {Eyink}, Physical Review Letters {\bf 104},  081101
  (2010).

\bibitem[Armstrong et al(1995)]{armstrong1995}
J.~W. {Armstrong}, B.~J. {Rickett}, and S.~R. {Spangler}, \apj {\bf 443},  209
  (1995).

\bibitem[Beresnyak \& Lazarian(2009)]{BL09a}
A. {Beresnyak} and A. {Lazarian}, \apj {\bf 702},  460  (2009).

\bibitem[Beresnyak et al(2009)]{BJL09}
A. {Beresnyak}, T.~W. {Jones}, and A. {Lazarian}, \apj {\bf 707},  1541
  (2009).

\bibitem[Beresnyak(2011)]{B11}
A. {Beresnyak}, Physical Review Letters {\bf 106},  075001  (2011).

\bibitem[Beresnyak \& Lazarian(2009b)]{BL09b}
A. {Beresnyak} and A. {Lazarian}, \apj {\bf 702},  1190  (2009).

\bibitem[Brandenburg \& Subramanian(2005)]{brandenburg2005}
A. {Brandenburg} and K. {Subramanian}, \physrep {\bf 417},  1  (2005).

\bibitem[Cho et al(2009)]{CVB09}
J. {Cho} {\it et~al.}, \apj {\bf 693},  1449  (2009).

\bibitem[Colgate et al.(2011)]{colgate} Colgate, S.~A. et al. 2011, Phys. Rev. Lett., 106, 175003 

\bibitem[Eyink(2005)]{eyink2005}
G.~L. {Eyink}, Physica D Nonlinear Phenomena {\bf 207},  91  (2005).

\bibitem[Eyink et al(2011)]{eyink2011}
G.~L. {Eyink}, A. {Lazarian}, and E.~T. {Vishniac}, ArXiv e-prints  (2011).


\bibitem[Eyink(2010)]{eyink2010}
G.~L. {Eyink}, \pre {\bf 82},  046314  (2010).


\bibitem[Goldreich \& Sridhar(1995)]{GS95}
Goldreich, P., \& Sridhar, S. 1995, ApJ, 438, 763

\bibitem[Goldstein et al(1995)]{goldstein1995}
B.~E. {Goldstein} {\it et~al.}, \grl {\bf 22},  3393  (1995).


\bibitem[Haugen et al(2004)]{haugen2004}
N.~E. {Haugen}, A. {Brandenburg}, and W. {Dobler}, \pre {\bf 70},  016308
  (2004).


\bibitem[K{\"a}pyl{\"a} et al(2009)]{kapyla2009}
P.~J. {K{\"a}pyl{\"a}}, M.~J. {Korpi}, and A. {Brandenburg}, \aap {\bf 500},
  633  (2009).

\bibitem[Kazantsev(1968)]{kazantsev1968}
A.~P. {Kazantsev}, Soviet JETP {\bf
  26},  1031  (1968).

\bibitem[Kulsrud \& Anderson(1992)]{kulsrud1992}
R.~M. {Kulsrud} and S.~W. {Anderson}, \apj {\bf 396},  606  (1992).

\bibitem[Laing et al(2008)]{Laing2008}
R.~A. {Laing}, A.~H. {Bridle}, P. {Parma}, and M. {Murgia}, \mnras {\bf 391},
  521  (2008).

\bibitem[Lazarian \& Vishniac(1999)]{lazarian1999}
A. {Lazarian} and E.~T. {Vishniac}, \apj {\bf 517},  700  (1999).

\bibitem[Ray et al(2011)]{ray2011}
S.~S. {Ray}, U. {Frisch}, S. {Nazarenko}, and T. {Matsumoto}, \pre {\bf 84},
  016301  (2011).

\bibitem[Ryu et al(2008)]{ryu2008}
D. {Ryu}, H. {Kang}, J. {Cho}, and S. {Das}, Science {\bf 320},  909  (2008).

\bibitem[Schekochihin \& Cowley(2007)]{schekochihin2007}
A.~A. {Schekochihin} and S.~C. {Cowley},  in {\em {Turbulence and Magnetic
  Fields in Astrophysical Plasmas}} (Springer, ADDRESS, 2007), pp.\ 85--+.


\bibitem[Schekochihin et al(2004)]{Schekochihin2004}
A.~A. {Schekochihin} {\it et~al.}, \apj {\bf 612},  276  (2004).

\bibitem[Sur et al.(2010)]{sur} Sur, S., Schleicher, 
D.~R.~G., Banerjee, R., Federrath, C., 
\& Klessen, R.~S.\ 2010, \apjl, 721, L134


\bibitem[Vainshtein \& Zeldovich(1972)]{vainshtein1972}
S. Vainshtein and Y. Zeldovich, Physics-Uspekhi {\bf 15},  159  (1972).

\bibitem[Vishniac \& Cho(2001)]{vishniac2001}
E.~T. {Vishniac} and J. {Cho}, \apj {\bf 550},  752  (2001).

\end{thebibliography}
\end{document}